\documentclass[prl,twocolumn,letterpaper,showpacs,groupedaddress,superscriptaddress,nofootinbib,floatfix,preprintnumbers,tightenlines]{revtex4-1}
\usepackage[bookmarks=false]{hyperref}
\usepackage{amssymb,amsmath,graphicx}
\usepackage{bm}
\usepackage{color}
\usepackage[tight]{subfigure}

\def\si{^1 \hskip -0.03in S _0}
\def\siii{^3 \hskip -0.025in S _1}

\def\pislash{{\pi\hskip-0.55em /}}

\def\L1Abar{\overline{L}_{1,A}}

\begin{document}

\begin{figure}[!t]
\vskip -1.1cm
\leftline{
\includegraphics[width=3.0 cm]{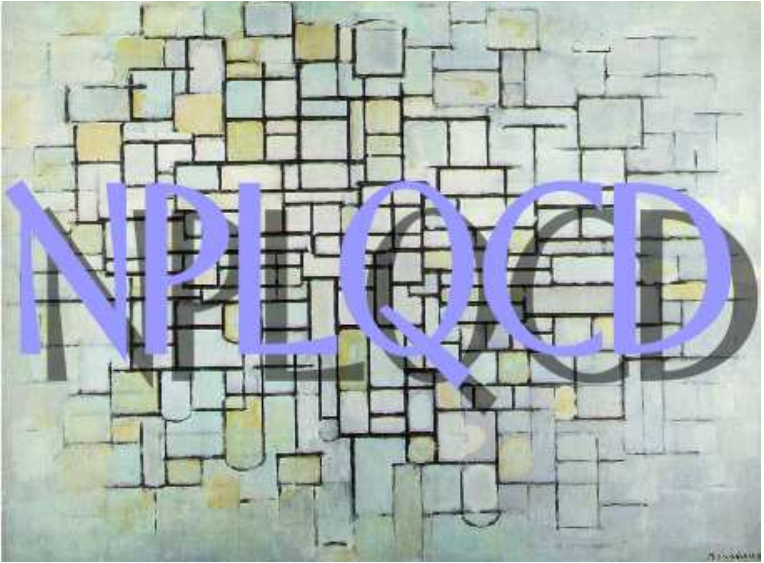}}
\vskip -0.5cm
\end{figure}

\title{Proton-proton fusion and tritium $\beta$-decay from lattice quantum chromodynamics}

\author{Martin J. Savage}
\affiliation{Institute for Nuclear Theory, University of Washington, Seattle, WA 98195-1550, USA}
\affiliation{Kavli Institute for Theoretical Physics, University of California, Santa Barbara, CA 93106, USA}

 \author{Phiala E. Shanahan } \affiliation{
 	Center for Theoretical Physics, 
 	Massachusetts Institute of Technology, 
 	Cambridge, MA 02139, USA}
\affiliation{Kavli Institute for Theoretical Physics, University of California, Santa Barbara, CA 93106, USA}

 \author{Brian C. Tiburzi} 
\affiliation{ Department of Physics, The City College of New York, New York, NY 10031, USA }
\affiliation{Graduate School and University Center, The City University of New York, New York, NY 10016, USA }
\affiliation{RIKEN BNL Research Center, Brookhaven National Laboratory, Upton, NY 11973, USA }
\affiliation{Kavli Institute for Theoretical Physics, University of California, Santa Barbara, CA 93106, USA}

\author{Michael L. Wagman} 
\affiliation{Department of Physics,
	University of Washington, Box 351560, Seattle, WA 98195, USA}
\affiliation{Kavli Institute for Theoretical Physics, University of California, Santa Barbara, CA 93106, USA}

\author{Frank Winter}
\affiliation{Jefferson Laboratory, 12000 Jefferson Avenue, 
	Newport News, VA 23606, USA}

\author{Silas~R.~Beane} 
\affiliation{Department of Physics,
	University of Washington, Box 351560, Seattle, WA 98195, USA}
\affiliation{Kavli Institute for Theoretical Physics, University of California, Santa Barbara, CA 93106, USA}

\author{Emmanuel~Chang}
\affiliation{Institute for Nuclear Theory, University of Washington, Seattle, WA 98195-1550, USA}

 \author{Zohreh Davoudi} \affiliation{
 	Center for Theoretical Physics, 
 	Massachusetts Institute of Technology, 
 	Cambridge, MA 02139, USA}
\affiliation{Kavli Institute for Theoretical Physics, University of California, Santa Barbara, CA 93106, USA}

 \author{William Detmold} \affiliation{
 	Center for Theoretical Physics, 
 	Massachusetts Institute of Technology, 
 	Cambridge, MA 02139, USA}
 \affiliation{Kavli Institute for Theoretical Physics, University of California, Santa Barbara, CA 93106, USA}
 
 \author{Kostas~Orginos}
 \affiliation{Department of Physics, College of William and Mary, Williamsburg,
 	VA 23187-8795, USA}
 \affiliation{Jefferson Laboratory, 12000 Jefferson Avenue, 
 	Newport News, VA 23606, USA}

\collaboration{NPLQCD Collaboration}

\date{\today}

\preprint{INT-PUB-16-033}
\preprint{JLAB-THY-16-2362}
\preprint{NSF-KITP-16-151}
\preprint{MIT-CTP-4844}
\preprint{NT@UW-16-12}

\pacs{11.15.Ha, 
      12.38.Gc, 
      13.40.Gp  
}

\begin{abstract}
The  nuclear matrix element determining the
$pp\to d e^+ \nu$ fusion cross section and the Gamow-Teller matrix element contributing to tritium $\beta$-decay
are calculated with lattice Quantum Chromodynamics (QCD) for the first time.
Using a new implementation of the background field method, these quantities are calculated 
at the SU(3)-flavor--symmetric value 
of the quark masses, corresponding to a pion mass of $m_\pi\sim 806$~MeV.
The   Gamow-Teller matrix element in tritium is found to be 
{$0.979(03)(10)$} at these quark masses, which is within $2\sigma$ of the 
experimental value.
Assuming that the short-distance correlated two-nucleon contributions to the 
matrix element (meson-exchange currents) depend only mildly on the quark masses, as seen for the analogous 
magnetic interactions, the calculated  $pp\to d e^+ \nu$ transition matrix element leads to a fusion cross section at the physical quark masses that is consistent with its currently accepted value. Moreover, the leading two-nucleon axial counterterm of pionless effective field theory is determined to be 
{ $L_{1,A}= 3.9(0.2)(1.0)(0.4)(0.9) ~{\rm fm}^3$}
at a renormalization scale set by the physical pion mass, also  agreeing within the accepted phenomenological range. 
This work concretely demonstrates that weak transition amplitudes in few-nucleon systems can be studied directly from the fundamental quark and gluon degrees of freedom and opens the way for subsequent investigations of many important quantities in nuclear physics.
\end{abstract}

\maketitle


Weak nuclear processes play a central role in many settings, from the instability of the neutron to the dynamics of core-collapse supernova. In this work,  the results of the first lattice Quantum Chromodynamics (LQCD) calculations of two such processes are presented, namely the $pp\to d e^+ \nu_e$ fusion process and  tritium $\beta$-decay.
The   $pp\to d e^+ \nu$ process is centrally important in astrophysics as it is primarily responsible for initiating the proton-proton fusion chain reaction
that provides the dominant energy production mechanism in stars like the Sun.
Significant theoretical effort has been expended in refining calculations of the $pp\to d e^+ \nu$  cross section 
at the energies relevant to solar burning, and progress continues to be made with a range of techniques~\cite{Schiavilla:1998je,Kong:1999tw,Kong:2000px,Butler:2001jj,Baroni:2015uza,Baroni:2016xll,Park:1993jf,Park:2000ct,Park:2002yp,Ando:2008va},
as summarized in Ref.~\cite{Adelberger:2010qa}.
This process is related to the $\overline{\nu} d\to n n e^+$ neutrino-induced deuteron-breakup reaction~\cite{Butler:1999sv,Butler:2000zp,Butler:2002cw}, relevant to the 
measurement of neutrino oscillations at the Sudbury Neutrino Observatory (SNO) \cite{Ahmad:2002jz,Chen:2002pv}, and to the muon capture reaction, $\mu^- d\to nn\nu_\mu$, which is the focus of current investigation in the MuSun experiment \cite{Andreev:2010wd}.
The second process studied in this work, tritium $\beta$-decay, is a powerful tool for investigating the weak interactions of the Standard Model and plays an important role in the search for new physics. 
The super-allowed process
$^3{\rm H}\to {}^3{\rm He} \ e^- \bar{\nu}$ is 
theoretically clean and is the simplest 
semileptonic  weak decay 
of a nuclear system.
In contrast to $pp$ fusion, this decay has been  very precisely studied in the laboratory
(see Ref.~\cite{Otten:2008zz} for a review) and provides important constraints on the antineutrino mass~\cite{Drexlin:2013lha}.
Tritium $\beta$-decay is also potentially sensitive to sterile neutrinos \cite{Mertens:2014nha,Barry:2014ika} 
and to interactions not present in the 
Standard Model~\cite{Stephenson:2000mw,Bonn:2007su,Barry:2014ika,Ludl:2016ane}.
Although the dominant contributions to the decay rate are under 
 theoretical control as this is a super-allowed process, the Gamow-Teller (GT) contribution (axial current) is somewhat 
more challenging to address than the Fermi (F) contribution (vector current).  
Improved constraints on multi-body contributions to the GT matrix element will translate into reduced uncertainties in predictions for decay rates of larger nuclei and are a first step towards understanding the quenching of $g_A$ in nuclei \cite{Buck:1975ae,Krofcheck:1985fg,Chou:1993zz}, a long-standing problem in nuclear theory.

In this Letter, LQCD is used to study the $pp\to d e^+ \nu_e$ fusion process and the
Gamow-Teller matrix element contributing to tritium $\beta$-decay for the first time, albeit at unphysically large values of the light quark masses and neglecting the effects of isospin-breaking and electromagnetism. 
This is accomplished using a new algorithm for implementing background fields, which is superior to existing methods. 
Further, the quantities of interest are extracted at high precision using a refined analysis strategy made possible by this new approach. 
For  $pp\to d e^+ \nu_e$, the deviations from the single-nucleon contributions are small but are well resolved with the new technique.
The leading  two-nucleon axial counterterm of pionless effective field theory ($\pislash$EFT), $L_{1,A}$, is determined. 
The axial coupling of $^3$H that determines the matrix element for $^3{\rm H}\to {}^3{\rm He}\ e^- \bar{\nu}$ 
in the isospin limit is found to be slightly smaller than that of the proton and is
consistent with previous phenomenological estimates \cite{Baroni:2016xll}.

\vspace*{3mm}
{\it Background Axial Fields:}
Background field techniques were first used in LQCD in the pioneering works of 
Ref.~\cite{Fucito:1982ff} and Refs.~\cite{Martinelli:1982cb,Bernard:1982yu} 
in the cases of axial and magnetic fields, respectively. 
Significant effort has been applied to using background electromagnetic fields to extract 
magnetic moments and electromagnetic polarizabilities of 
hadrons~\cite{Lee:2005dq,Detmold:2006vu,Detmold:2009dx,Detmold:2010ts,Parreno:2016fwu} 
and nuclei \cite{Beane:2014ora,Chang:2015qxa,Detmold:2015daa}, as well as the magnetic transition amplitude for the $n p \to d \gamma$ process~\cite{Beane:2015yha}. Very recently, axial background fields have been employed to extract the axial charge of the proton~\cite{Chambers:2014qaa,Chambers:2015bka}, and generalizations to nonzero momentum transfer~\cite{Detmold:2004kw,Davoudi:2015cba,Bali:2015msa} have been used~\cite{Chambers:2015kuw} to access the axial form factor of the nucleon.

In this work, a new method is used to generate hadronic correlation functions 
order-by-order in the background field. 
In the standard approach, correlation functions are constructed from the contraction of quark propagators
that are modified by the presence of a background field.  
The same effect can be achieved by directly constructing propagators that include explicit current insertions, then using such propagators to construct correlation functions. 
For the quantities studied in this work only a single insertion of the background axial field is required. 
To this end, the {\it compound propagator}
\begin{eqnarray}
S^{(q)}_{\lambda_q;\Gamma}(x,y) & = & S^{(q)}(x,y)
+\lambda_q \int \! dz\ S^{(q)}(x,z) \Gamma  S^{(q)}(z,y)\ \ \ \ 
\label{eq:bfprop}
\end{eqnarray}
is constructed for $\Gamma=\gamma_3\gamma_5$ and flavors $q=\{u,d\}$, where 
$S^{(q)}(x,y)$ is the quark propagator of flavor $q$ and $\lambda_q$ is a constant (a similar approach is implemented in Ref.~\cite{deDivitiis:2012vs} in a different context). 
The second term in this expression is computed using standard sequential source techniques and the procedure can be repeated to produce propagators with higher-order couplings. 
These compound propagators are sufficient to construct the isovector axial matrix elements for zero momentum insertion in any hadronic or nuclear system  (isoscalar responses, which also involve insertions on the sea-quark propagators, are not addressed).
This work focuses on zero-momentum--projected correlation functions,
\begin{eqnarray}
C^{(h)}_{\lambda_u;\lambda_d}(t) 
& = & 
\sum_{\bf x}
\langle 0| \chi_h({\bf x},t) \chi^\dagger_h(0) |0 \rangle_{\lambda_u;\lambda_d},
\label{eq:bfcorr}
\end{eqnarray}
where $\langle\ldots\rangle_{\lambda_u;\lambda_d}$ denotes the expectation value determined using the compound propagators.
The interpolating operators for hadrons and nuclei, $\chi_h$, are those previously 
used to study spectroscopy of these systems~\cite{Beane:2012vq,Beane:2013br}.
By construction, $C_{\lambda_u;\lambda_d}^{(h)}(t)$ is a polynomial of maximum order 
$\lambda_u^{N_u}\lambda_d^{N_d}$ in the field strengths, where $N_{u(d)}$ is the number of up (down) 
quarks in the particular interpolating operator.

\vspace*{3mm}
{\it Details of the LQCD Calculation:}
The calculations presented below used an ensemble of gauge-field configurations generated with a clover-improved fermion action~\cite{Sheikholeslami:1985ij} 
and a L\"uscher-Weisz gauge action~\cite{Luscher:1984xn}. 
The ensemble was generated with $N_f=3$ degenerate light-quark flavors with masses tuned to the physical strange 
quark mass, producing a pion of mass $m_\pi\sim 806~{\rm MeV}$, 
with a volume of $L^3\times T=32^3\times48$ and a lattice spacing of $a\sim 0.145~{\rm fm}$ (as determined from $\Upsilon$ spectroscopy).
For these calculations, 437 configurations, with a spacing of 10 trajectories between configurations, were used.
Correlation functions  were computed for  
$h=\{p,n,d,nn,np(\si),pp,{}^3{\rm H},{}^3{\rm He}\}$ from propagators generated from a smeared source 
and either a smeared (SS) or point (SP) sink. Sixteen 
different source locations were averaged over on each configuration. 
Compound propagators and correlation functions were calculated at six different values of the background field strength parameter 
$\lambda=\{\pm 0.05, \pm 0.1, \pm 0.2\}$.
The axial current renormalization factor 
{$Z_A=0.867(43)$} was determined from computations of the vector current in the proton, noting that $Z_A=Z_V + {\cal O}(a)$ and assigning a $5\%$ systematic 
uncertainty associated with lattice-spacing artifacts (statistical uncertainties are negligible). A determination that removes the leading lattice-spacing artifacts leads to $Z_A=0.8623(01)(71)$~\cite{GreenLatt2016,Yoon:2016jzj}
at a pion mass of $m_\pi\sim 317~{\rm MeV}$.

\begin{figure}[!t]
	\includegraphics[width=\columnwidth]{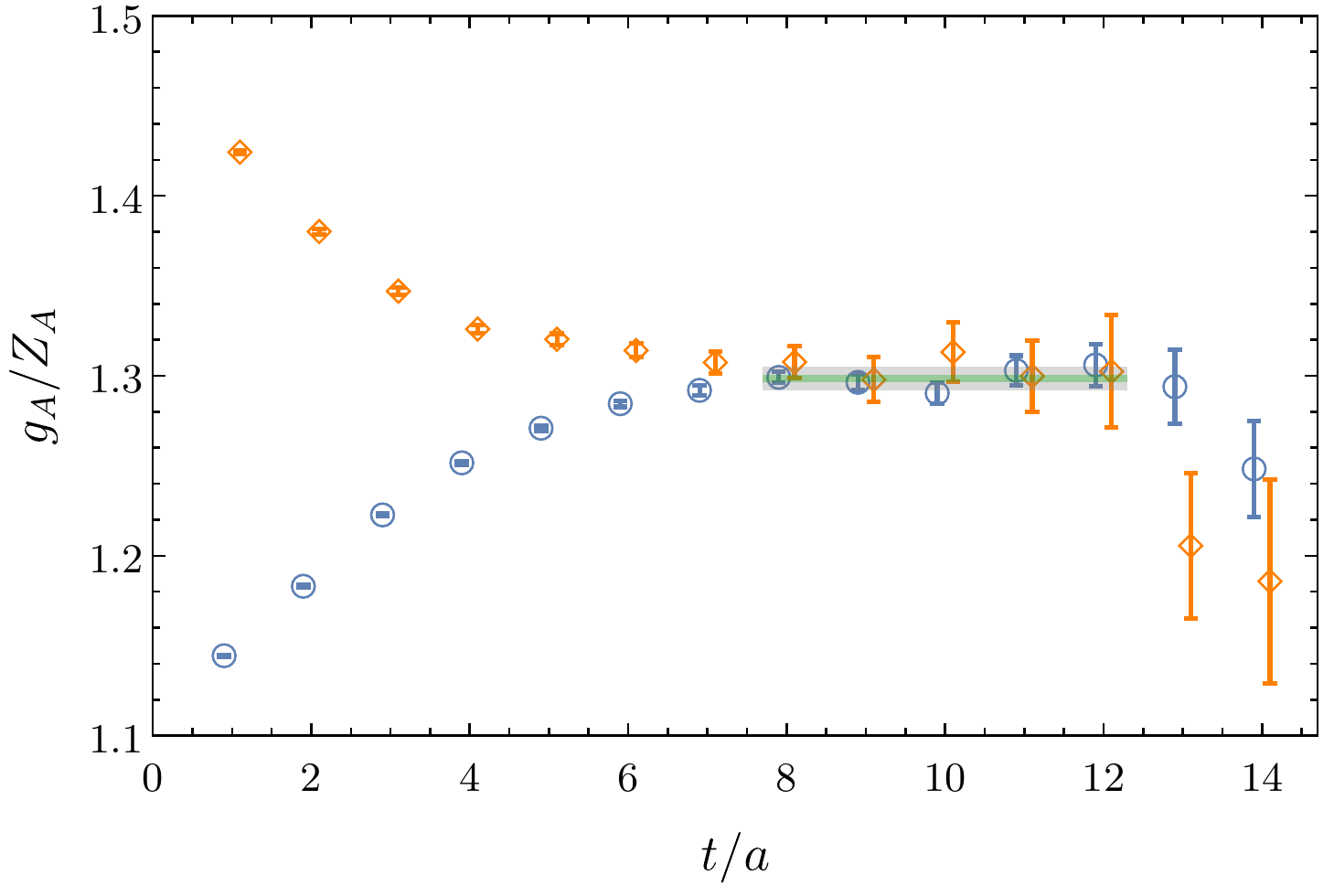}
	\caption{\label{fig:gA} 
The ratios of correlation functions that determine the unrenormalized isovector axial charge of the proton. The orange diamonds (blue circles) correspond to the SS (SP)  
effective correlator ratios, $\overline{R}_p(t)$,  defined in Eq.~(\ref{eq:gAeff}), and the band corresponds to a constant fit to the plateau interval of both SS and SP.
}
\end{figure}

\vspace*{3mm}
{\it The Proton Axial Charge:}
The simplest matrix element of the isovector axial current determines the axial charge of the proton. 
The correlation function $C^{(p)}_{\lambda_u;\lambda_d}(t)$ is at most quadratic in $\lambda_u$ and linear in $\lambda_d$ 
when constructed from the compound propagators 
$S^{(u)}_{\lambda_{u};\gamma_3\gamma_5}(x,y)$ and $S^{(d)}_{\lambda_{d};\gamma_3\gamma_5}(x,y) $, 
as the proton has two valence
up quarks and one valence down quark. 
Consequently, using at least one (two) nonzero value(s) of $\lambda_{d(u)}$ enables extraction of the axial current matrix element as the linear response by using suitable polynomial fits.
The difference of the up-quark and down-quark matrix elements
can be used to construct the desired three-point function 
containing the isovector axial current. This can then be combined with the zero-field two-point function
to form a ratio that asymptotes to the desired axial charge at late times, namely
\begin{eqnarray}
R_p(t)
&=&
\frac{
\left.
C^{(p)}_{\lambda_u;\lambda_d=0}(t)\right|_{{\cal O}(\lambda_u)} 
- \left. C^{(p)}_{\lambda_u=0;\lambda_d}(t)
\right|_{{\cal O}(\lambda_d)}
}
{ C^{(p)}_{\lambda_u=0;\lambda_d=0}(t)},
\label{eq:Rdef}
\end{eqnarray}
where the ratios are spin-weighted averages, and 
``$\big|_{{\cal O}(\lambda_q)}$'' extracts the coefficient of $\lambda_q$ 
in the preceding expression. 
Then, 
\begin{eqnarray}
\overline{R}_p(t) \equiv R_p(t+1)-R_p(t) &\stackrel{t \to \infty }{\longrightarrow}& \frac{g_A}{Z_A}
\, ,
\label{eq:gAeff}
\end{eqnarray}
{
where corrections to this relation 
from backwards propagating states originating from the finite extent of the time direction
are suppressed by at least $e^{-2m_\pi T/3}\sim 10^{-7}$ in the signal region in the present set of calculations.}
The {\it effective-$g_A$ plots} 
resulting from the correlator differences are shown in Fig.~\ref{fig:gA}, along with the result of a combined constant fit to the SS and SP ratios that extracts $g_A/Z_A$ from the late-time asymptote.
The extracted value is
{$g_A/Z_A = 1.298(2)(7)$},
where the first  uncertainty is statistical (determined from a bootstrap analysis) and the second is systematic 
(arising from choices of fit ranges in both the field strengths and temporal separation as well as from differences in analysis techniques).
Including the renormalization factor yields an axial charge of 
{$g_A = 1.13(2)(7)$}, which is consistent with previous determinations from standard three-point function techniques at this  pion mass~\cite{Yoon:2016dij,Lin:2011ab}.

\vspace*{3mm}
{\it The GT Matrix Element for Tritium $\beta$-decay:}
The half-life of tritium, $t_{1/2}$, is related to the F and GT matrix elements by~\cite{Schiavilla:1998je} 
\begin{eqnarray}
\frac{(1+\delta_R) f_V }{ K/G_V^2} t_{1/2} = \frac{1}{\langle {\bf F} \rangle^2 +f_A/f_V\,g_A^2\langle {\bf GT}\rangle^2},
\end{eqnarray}
where the factors on the left-hand side are known precisely from theory or experiment. On the right-hand side, 
$f_{A,V}$ denote known Fermi functions~\cite{Simpson:1987zz} and $\langle {\bf F} \rangle$ and 
$\langle {\bf GT} \rangle$ are the F and GT reduced matrix elements, respectively. 
The Ademollo-Gatto theorem \cite{PhysRevLett.13.264} implies $\langle {\bf F} \rangle\sim1$, modified only by 
second-order isospin-breaking and by electromagnetic corrections.
However,  
$
  \langle {}^3{\rm He}| \overline{q}\gamma_k \gamma_5 \tau^+ q | {}^3{\rm H}\rangle =
 \overline{u} \gamma_k \gamma_5 \tau^+ u \ g_A \langle{\bf GT}\rangle
 $ 
(assuming vanishing electron mass and at vanishing lepton momentum)
is less constrained, and its evaluation is the focus of this section.

\begin{figure}[!t]
	\includegraphics[width=\columnwidth]{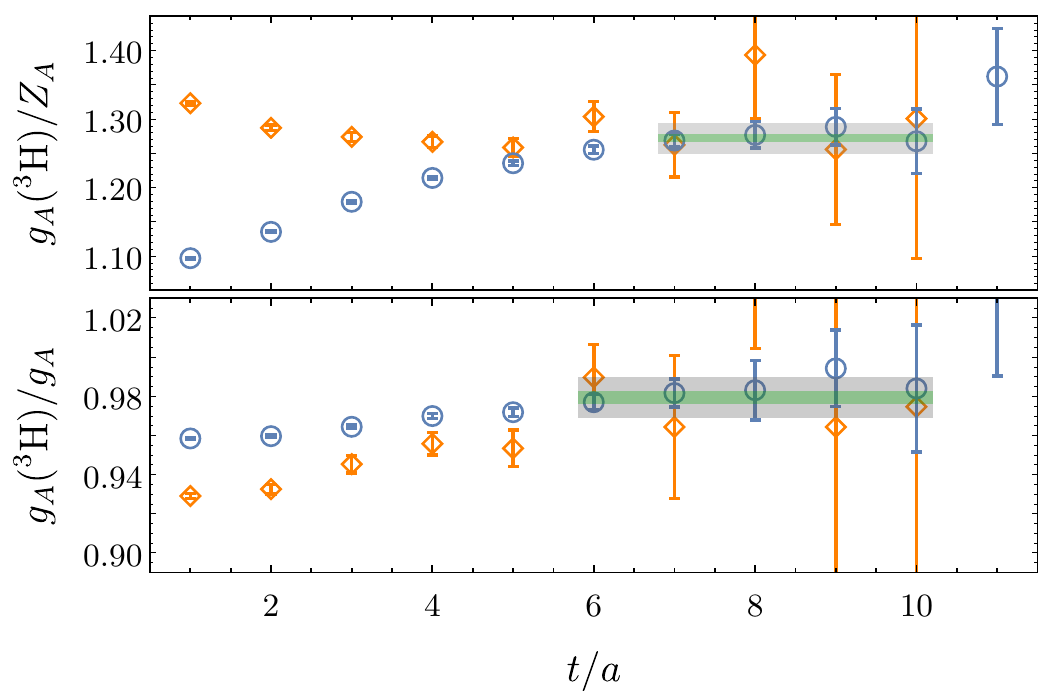}
	\caption{\label{fig:tritium} 
	The ratios of correlation functions that determine the unrenormalized isovector axial matrix element in $^3$H (upper panel),  
	and the ratio of the isovector axial matrix element in $^{3}$H to that in the proton (lower panel). 
	The orange diamonds (blue circles) correspond to the SS (SP) effective correlator ratios and 
	the bands correspond to constant fits to the asymptotic behavior.
	}		
\end{figure}

By isospin symmetry, the spin-averaged GT matrix element 
for $^3$H$\rightarrow ^3$He$\ e^-\overline{\nu}$
is related to the axial charge of the triton, $g_A({}^3{\rm H})$,
when the light quarks are degenerate and in the absence of electromagnetism. 
Analogous to $R_p(t)$ in Eq.~(\ref{eq:Rdef}), the ratio $R_{{}^3 \rm H}(t)$ of correlation 
functions in background fields is constructed such that, analogous to Eq.~(\ref{eq:gAeff}), $\overline{R}_{{}^3\rm H}(t)\to g_A({}^3{\rm H})/Z_A$ in the large-time limit. 
The analysis of these correlation functions is more complex 
than for the proton because the triton has four up quarks and 
five down quarks and the correlators are thus quartic and quintic polynomials in $\lambda_{u,d}$, respectively.
Polynomial fits to the calculated correlation functions 
are sufficient to extract the terms linear in $\lambda_{u,d}$.
Results for $\overline{R}_{{}^3H}(t)$ are shown in Fig.~\ref{fig:tritium} along with a constant fit to the asymptotic 
value $g_A({}^3{\rm H})/Z_A$.
Also shown in Fig.~\ref{fig:tritium} is $\langle{\bf GT}\rangle(t) = \overline{R}_{{}^3H}(t)/\overline{R}_p(t)$, which is independent of $Z_A$,
and the fit to its asymptotic value, $g_A({}^3{\rm H})/g_A$.
Analyses of these ratios lead to 
\begin{eqnarray}
\frac{g_A({}^3{\rm H})}{Z_A} = { 1.272(6)(22) },~ 
\frac{g_A({}^3{\rm H})}{g_A} =  { 0.979(3)(10) },
\label{eq:3H}
\end{eqnarray}
where the first uncertainties are statistical and the second arise from systematics as described for $g_A$.
The result for $g_A({}^3{\rm H})/g_A$ is quite close to the precise, experimentally-determined value of 
$\langle{\bf GT}\rangle = 0.9511(13)$~\cite{Baroni:2016xll}
at the physical quark masses.
In the context of $\pislash$EFT,
the short-distance two-nucleon axial-vector operator, with coefficient $L_{1,A}$~\cite{Butler:2001jj}, 
is expected to give the leading contribution to the difference of this ratio from unity \cite{De-Leon:2016wyu}.

\vspace*{3mm}
{\it The Low-Energy Proton-Proton Fusion Cross Section:}
The low-energy cross section for $pp\to d e^+ \nu$ is dictated by the matrix element
\begin{eqnarray}
\left|\left\langle d;j\left| A_k^- \right|pp\right\rangle\right|
\equiv
g_A C_\eta \sqrt\frac{32 \pi}{\gamma^3} \,\Lambda(p) \,\delta_{jk},
\label{eq:ppMATdef}
\end{eqnarray}
where $A_k^a(x)$ is the axial current with isospin and spin components $a$ and $k$ respectively, $j$ is the deuteron spin index, $C_\eta$ is the Sommerfeld factor and $\gamma$ is the deuteron binding momentum. The quantity $\Lambda(p)$ 
has been calculated at threshold in $\pislash$EFT to N$^2$LO~\cite{Kong:2000px} and N$^4$LO~\cite{Butler:2001jj} and later with a dibaryon approach \cite{Ando:2008va,De-Leon:2016wyu} and in pionful EFT \cite{Acharya:2016kfl}. With the approach of Ref.~\cite{Butler:2001jj}, resumming all of the effective range contributions \cite{Beane:2000fi, Phillips:1999hh, Ando:2008va}, $\Lambda(0)$ at N$^{2}$LO is related to the renormalization-scale independent short-distance quantity $L^{sd-2b}_{1,A}$ that is a solely two-body contribution, 
along with scattering parameters and Coulomb corrections:
\begin{eqnarray}
&&\Lambda(0) =
{1 \over\sqrt{1- \gamma \rho}}  
\{ e^\chi - \gamma a_{pp} [ 1-\chi e^\chi \Gamma(0,\chi) ]+~~~~~~~~
\nonumber\\
 &&\qquad
{1\over 2} \gamma^2 a_{pp} \sqrt{r_1 \rho} \}- {1\over 2g_A }\gamma a_{pp} \sqrt{1-\gamma \rho} \ L^{sd-2b}_{1,A}.~~
\label{eq:Lambda-0}
\end{eqnarray}
Here $\chi=\alpha M_p /\gamma$, where $\alpha$ is the QED fine-structure constant and $M_p$ is the mass of the proton.  The $pp$ scattering length is  $a_{pp}$,
$r_1$ and $\rho$ are the effective ranges in the $\si$ and $\siii$ channels, respectively,
and $\Gamma(0,\chi)$ is the incomplete gamma function.
 A determination of  $ L^{sd-2b}_{1,A}$, or equivalently of the  $\pislash$EFT coupling $L_{1,A}$
which is determined from the scale-independent constant
 \begin{eqnarray}
\overline{L}_{1,A} &=&
{1\over 2g_A} {1-\gamma \rho\over\gamma} \ L^{sd-2b}_{1,A}  - {1\over 2} \sqrt{r_1 \rho}
\end{eqnarray} 
as shown explicitly in Ref.~\cite{Butler:2001jj},
is a  goal of the present LQCD calculations.

\begin{figure}[!t]
	\includegraphics[width=\columnwidth]{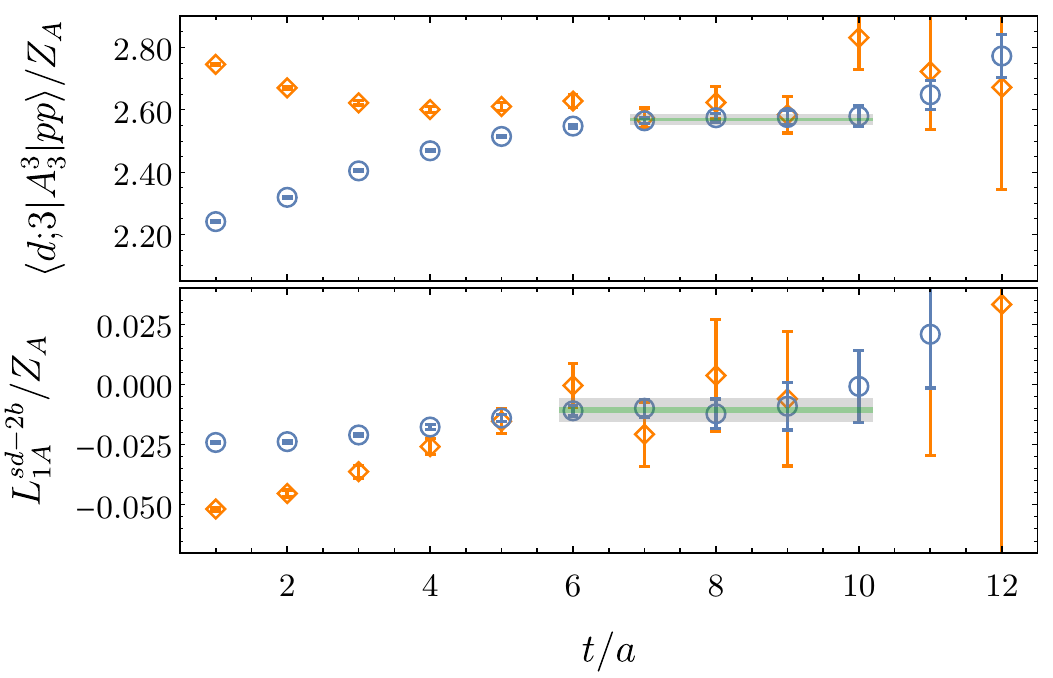}
	\caption{
		\label{fig:dtonp} 
The ratios of correlation functions that determine the unrenormalized isovector axial matrix element in the $J_z=I_z=0$ coupled two-nucleon system (upper panel), and the unrenormalized difference between the axial matrix element in this channel and $2g_A$ (lower panel). The orange diamonds (blue circles) correspond to the SS (SP) effective correlator ratios and the bands correspond to fits to the asymptotic plateau behavior {and include only the statistical and fitting systematic uncertainties (the additional 1\% uncertainty from Wigner symmetry breaking is not represented in the bands)}.}
\end{figure}

A background  isovector axial-vector  field mixes the $J_z~=~I_z~=0$ components of the 
$\siii$ and $\si$ two-nucleon channels, enabling the 
$pp$-fusion matrix element to be accessed.
Using the new background field construction, the relevant off-diagonal matrix element $C_{\lambda_u;\lambda_d}^{({\siii},{\si})}(t)$ is a cubic polynomial in both $\lambda_{u}$ and $\lambda_{d}$. 
In Ref.~\cite{Beane:2015yha}, the analogous mixing between the two-nucleon channels induced by 
an isovector magnetic field 
was treated by diagonalizing a (channel-space) matrix of correlators and determining 
the splittings between energy eigenvalues. This provided access to the matrix element dictating $np\to d\gamma$ at low energies, as was proposed in Ref.~\cite{Detmold:2004qn}.
Such a method can also be used for the axial field, but the improved approach implemented here makes use of the finite-order 
polynomial structure to access the matrix element directly. For a background field that couples to the $u$ quarks,
\begin{eqnarray}
&& C_{\lambda_u;\lambda_d=0}^{(\siii,\si)}(t) 
=\ 
\lambda_u \sum_{\tau=0}^t \sum_{{\bf x},{\bf y}}\langle 0| \chi^3_{{}^3S_1}\!({\bf x},t) A_3^u({\bf y},\tau)\chi^\dagger_{{}^1S_0}\!(0) |0 \rangle
\nonumber \\
&  & \qquad\qquad\qquad\qquad\qquad\qquad\ +\ 
c_2 \lambda_u^2 + c_3\lambda_u^3,
\end{eqnarray}
where $\chi^3_{\siii}$ ($\chi_{\si}$) is an interpolating field for the $J_z=0$ ($I_z=0$) component of the $\siii$ ($\si$) channel, 
$A_3^u=\overline{u}\gamma_3\gamma_5 u$, and $c_{2,3}$ are irrelevant terms. 
Calculations of the background field correlators at three or more values of $\lambda_u$ allow for the extraction of the  
term that is linear in $\lambda_u$. A similar procedure yields the  
term that is linear in $\lambda_d$ from background fields  coupling to the d quark.
Taking the difference of the ratios of these terms to the corresponding  zero-field two-point functions determines the transition matrix element in the finite lattice volume;
\begin{eqnarray}
R_{\siii,\si}(t)&=&\frac{\left . C^{(\siii,\si)}_{\lambda_{u},\lambda_{d}=0}(t)\right|_{{\cal O}(\lambda_u)}
	-
	\left. C^{(\siii,\si)}_{\lambda_{u}=0,\lambda_{d}}(t)\right|_{{\cal O}(\lambda_d)} }{\sqrt{C_{\lambda_{u}=0,\lambda_{d}=0}^{(\siii,\siii)}(t)C_{\lambda_{u}=0,\lambda_{d}=0}^{(\si,\si)}(t)}}.\ \ \ \ \ 
\end{eqnarray}
Consequently, the difference between ratios at neighboring timeslices determines the isovector matrix element;
\begin{eqnarray}
\overline{R}_{\siii,\si}(t) &\equiv& R_{\siii,\si}(t+1)-R_{\siii,\si}(t)
\nonumber\\
&\stackrel{t \to \infty }{\longrightarrow}&   \frac{\left\langle \siii;J_z=0 \left| A_3^3 \right| {\si};I_z=0 \right\rangle}{Z_A}, 
\end{eqnarray}
in the limit where $\Delta E= E_d -E_{pp}$ is small (as is the case with the quark masses used in this calculation \cite{Beane:2012vq}), and when the contributions from excited states are suppressed. 
This quantity, measured with both SS and SP correlators, is shown in Fig.~\ref{fig:dtonp}, 
along with the extracted value of the axial matrix element, 
{ 
$\left\langle \siii;J_z=0 \left| A_3^3 \right| {\si};I_z=0 \right\rangle/Z_A = { 2.568(5)(31)}$,
where the first uncertainty is statistical and the 
second is a systematic encompassing choices of fit ranges in time, field strength
and variations in analysis techniques.  The latter includes an estimate 
of the violation of Wigner's SU(4) symmetry, contributing an uncertainty
${\cal O}\left(1/N_c^4\right)\sim 1\%$ to the extraction of the matrix element based on the large-$N_c$ limit.}
At the pion mass of this study, the initial and final two-nucleon states are deeply bound \cite{Beane:2012vq}
and the finite-volume effects in the matrix elements are negligible \cite{Briceno:2012yi,Briceno:2015tza}. At lighter values of the quark masses, where the $np(\si)$ system and/or the deuteron are unbound or only weakly bound, the connection between finite-volume matrix elements and transition amplitudes requires the framework developed in Refs.~\cite{Briceno:2012yi,Briceno:2015tza}.

To isolate the two-body contribution, the combination $L^{sd-2b}_{1,A}(t)/Z_A=[\overline{R}_{\siii,\si}(t)- 2\overline{R}_{p}(t)]/2$ is formed as shown in the lower panel of Fig.~\ref{fig:dtonp}. 
Taking advantage of the near-degeneracy of the  $^3S_1$ and $^1S_0$ two-nucleon channels at the quark masses used in this calculation, it is straightforward to show that this correlated difference leads directly to the short-distance two-nucleon quantity, $ L^{sd-2b}_{1,A}$. 
{ 
Fitting a constant to the late-time behavior of this quantity leads to
\begin{eqnarray}
\frac{L^{sd-2b}_{1,A}}{Z_A}  & = & \frac{ \left\langle {\siii};J_z=0 \left| A_3^3 \right| {\si};I_z=0 \right\rangle - 2 g_A}{2Z_A}
\nonumber\\
& = &  
-0.011(01) (15)  
\  ,
\end{eqnarray}
where the first uncertainty is statistical and the second encompasses fitting and analysis systematics.
}
 
In light of the 
mild quark-mass dependence of the analogous short-distance, two-body quantity contributing to 
$np\rightarrow d\gamma$~\cite{Beane:2015yha}, $L^{sd-2b}_{1,A}$
is 
likely
 to be largely insensitive to the 
 {pion mass between $m_\pi \sim 806$~MeV  and its physical value.
 This approximate independence and the associated systematic uncertainty will
 need to be refined in subsequent calculations.}
 Based on this expectation, the result obtained here at $m_\pi \sim 806$~MeV 
{is}
used to estimate the value of $L^{sd-2b}_{1,A}$ at the physical pion mass by including an additional 50\% additive uncertainty. Propagating this uncertainty through Eq.~(\ref{eq:Lambda-0}), the 
threshold value of $\Lambda(p)$ in this system 
at the physical quark masses is determined 
to be 
{$\Lambda(0)  =   2.659(2)(9)(5) $}, 
where the uncertainties are statistical, fitting and analysis systematic, and quark-mass extrapolation systematic, respectively. Uncertainties in the scattering parameters and other physical mass inputs are also propagated and included in the systematic uncertainty.
This result is remarkably close to the currently accepted, precise phenomenological value, $\Lambda(0) =2.65(1)$ \cite{Adelberger:2010qa} (see also Ref.~\cite{De-Leon:2016wyu}). 
The N$^2$LO relation of Ref.~\cite{Butler:2001jj}, when enhanced by the summation of the effective ranges to all orders using the dibaryon field approach \cite{Beane:2000fi, Phillips:1999hh, Ando:2008va}, gives $\Lambda(0)= {2.62(1)+0.0105(1) L_{1,A}}$, 
enabling a determination of the $\pislash$EFT coupling 
{ 
\begin{eqnarray}
L_{1,A}= 3.9(0.2)(1.0)(0.4)(0.9) ~{\rm fm}^3 , 
\end{eqnarray}
}
at a renormalization scale $\mu=m_{\pi}$. The uncertainties are statistical, fitting and analysis systematic, mass extrapolation systematic, and a power-counting estimate of higher order corrections in $\pislash$EFT, respectively.
This value is also very close to  previous phenomenological estimates, as summarized in Refs.~\cite{Butler:2002cw,Adelberger:2010qa}.

\vspace*{3mm}
{\it Summary:}
The primary results of this work are the isovector axial-current matrix elements in two and three-nucleon systems calculated directly from the underlying theory of the strong interactions using lattice QCD.\footnote{See Supplemental Material for additional discussion of technical aspects of these calculations, which includes Refs. \cite{comment,Berkowitz:2015eaa,Tiburzi:2017iux,Chen:1999tn,Iritani:2017rlk}.} These matrix elements determine the cross section for the $pp$ fusion process $pp\rightarrow d e^+\nu$
and the Gamow-Teller contribution to tritium $\beta$-decay, 
${}^3{\rm H}\rightarrow {}^3{\rm He}\ e^-\overline{\nu}$. 
While the calculations are performed at unphysical quark masses corresponding to $m_\pi \sim 806$ MeV and at a single lattice spacing and volume, 
the  mild mass dependence of the analogous short-distance quantity in the $np\to d\gamma$ magnetic transition enables  an estimate of 
the $pp\to de^+\nu$ matrix element at the physical values of the quark masses, and the results are found to agree within uncertainties with phenomenology.  
Future LQCD calculations, including electromagnetism beyond Coulomb effects, at lighter quark masses with isospin splittings, larger volumes, and finer lattice spacings, making use of the new techniques that are introduced here, will enable extractions of these axial matrix elements with fully quantified uncertainties and will be important for phenomenology, providing increasingly precise values for the $pp$-fusion cross section and GT matrix element in tritium $\beta$ decay.

Beyond the current study, background axial-field calculations also allow the extraction of 
second-order, as well as momentum-dependent, responses to axial fields. 
Second-order responses are  
important for determining nuclear $\beta\beta$-decay matrix elements, both with and without (for a light Majorana neutrino) the emission of associated neutrinos \cite{Shanahan:2017bgi}.  
Momentum-dependent axial background fields will allow the determination of nuclear effects in neutrino-nucleus scattering.   
In both cases, LQCD calculations of these quantities in light nuclei will provide vital input with which to constrain the 
nuclear many-body methods that are used to determine the matrix elements for these processes in heavy nuclei.

\vspace*{2mm}

{\it Acknowledgments:}
	We would like to thank Jiunn-Wei Chen and Peter Kammel for several interesting discussions. 
	This research was supported in part by the National Science Foundation under grant number NSF PHY11-25915 and
	SRB, ZD, WD, MJS, PES, BCT and MLW acknowledge the Kavli Institute for Theoretical Physics for hospitality 
	during completion of this work.
	Calculations were performed using computational resources provided
	by  NERSC (supported by U.S. Department of
	Energy grant number DE-AC02-05CH11231),
	and by the USQCD
	collaboration.  This research used resources of the Oak Ridge Leadership 
	Computing Facility at the Oak Ridge National Laboratory, which is supported 
	by the Office of Science of the U.S. Department of Energy under Contract 
	number DE-AC05-00OR22725. 
	The PRACE Research Infrastructure resources  at the 
	Tr\`es Grand Centre de Calcul and Barcelona Supercomputing Center were also used.
	Parts of the calculations used the Chroma software
	suite~\cite{Edwards:2004sx}.  
	SRB was partially supported by NSF continuing grant number PHY1206498 and by the U.S. Department of Energy through grant
	number DE-SC001347.  
	EC was supported in part by the USQCD SciDAC project, the U.S. Department of Energy through 
	grant number DE-SC00-10337,  and by U.S. Department of Energy grant number DE-FG02-00ER41132.
	ZD, WD and PES were partly supported by  U.S. Department of Energy Early Career Research Award DE-SC0010495 and grant number DE-SC0011090.
	KO was partially supported by the U.S. Department of Energy through grant
	number DE- FG02-04ER41302 and through contract number DE-AC05-06OR23177
	under which JSA operates the Thomas Jefferson National Accelerator Facility.  
	MJS was supported  by DOE grant number~DE-FG02-00ER41132, and  in part by the USQCD SciDAC project, 
	the U.S. Department of Energy through grant number DE-SC00-10337.	
	BCT was supported in part by a joint City College of New York-RIKEN Brookhaven Research Center
	fellowship, and by the U.S. National Science Foundation, under grant
	number PHY15-15738. 
	MLW was supported  in part by DOE grant number~DE-FG02-00ER41132.
	FW was partially supported through the USQCD Scientific Discovery through Advanced Computing (SciDAC) project 
	funded by U.S. Department of Energy, Office of Science, Offices of Advanced Scientific Computing Research, 
	Nuclear Physics and High Energy Physics and by the U.S. Department of Energy, Office of Science, Office of Nuclear Physics under contract DE-AC05-06OR23177.

%

\bibliography{tritium.bib}

\newpage

\begin{widetext}
\section*{Supplementary Material}

In this section, supporting supplemental material is provided.

\subsection*{Correlation Functions}

The correlation functions that are generated and analyzed to determine  matrix elements of the isovector axial current in the 
proton are of the form,
\begin{eqnarray}
C^{(p\uparrow)}_{\lambda_u;\lambda_d=0}(t) 
&=&
\sum_{\bm x}^{}
\left(\ 
\vphantom{\sum_{\bm x}^{} }
\langle 0| \chi_{p\uparrow}({\bm x},t) \chi^\dagger_{p\uparrow}(0) |0 \rangle 
+ \lambda_u
\sum_{\bm y}\sum_{t_1=0}^t
\langle 0| \chi_{p\uparrow}({\bm x},t) J_3^{(u)} ({\bm y},t_1)  \chi^\dagger_{p\uparrow}(0) |0 \rangle \ \right)
+ d_2 \lambda_u^2, 
\nonumber 
\qquad \\
C^{(p\uparrow)}_{\lambda_u=0;\lambda_d}(t) 
& =& 
\sum_{\bm x}
\left(\ 
\vphantom{\sum_{\bm x}^{} }
\langle 0| \chi_{p\uparrow}({\bm x},t) \chi^\dagger_{p\uparrow}(0) |0 \rangle 
+ \lambda_d 
\sum_{\bm y}\sum_{t_1=0}^t
\langle 0| \chi_{p\uparrow}({\bm x},t) J_3^{(d)} ({\bm y},t_1) \chi^\dagger_{p\uparrow}(0) |0 \rangle
\ \right),
\label{eq:prot}
\end{eqnarray}
where $d_2$ is a higher-order term whose value is not necessary for the present analysis.
Only components  linear in $\lambda_q$ are needed in order to 
isolate the matrix element of the axial current.
These components are easily determined, correlation function-by-correlation function,
from calculations in an appropriate number of background fields, 
examples of which are shown in Fig.~\ref{fig:field_response}.
\begin{figure}[!h]
	\subfigure[]{
		\includegraphics[width=0.45\textwidth]{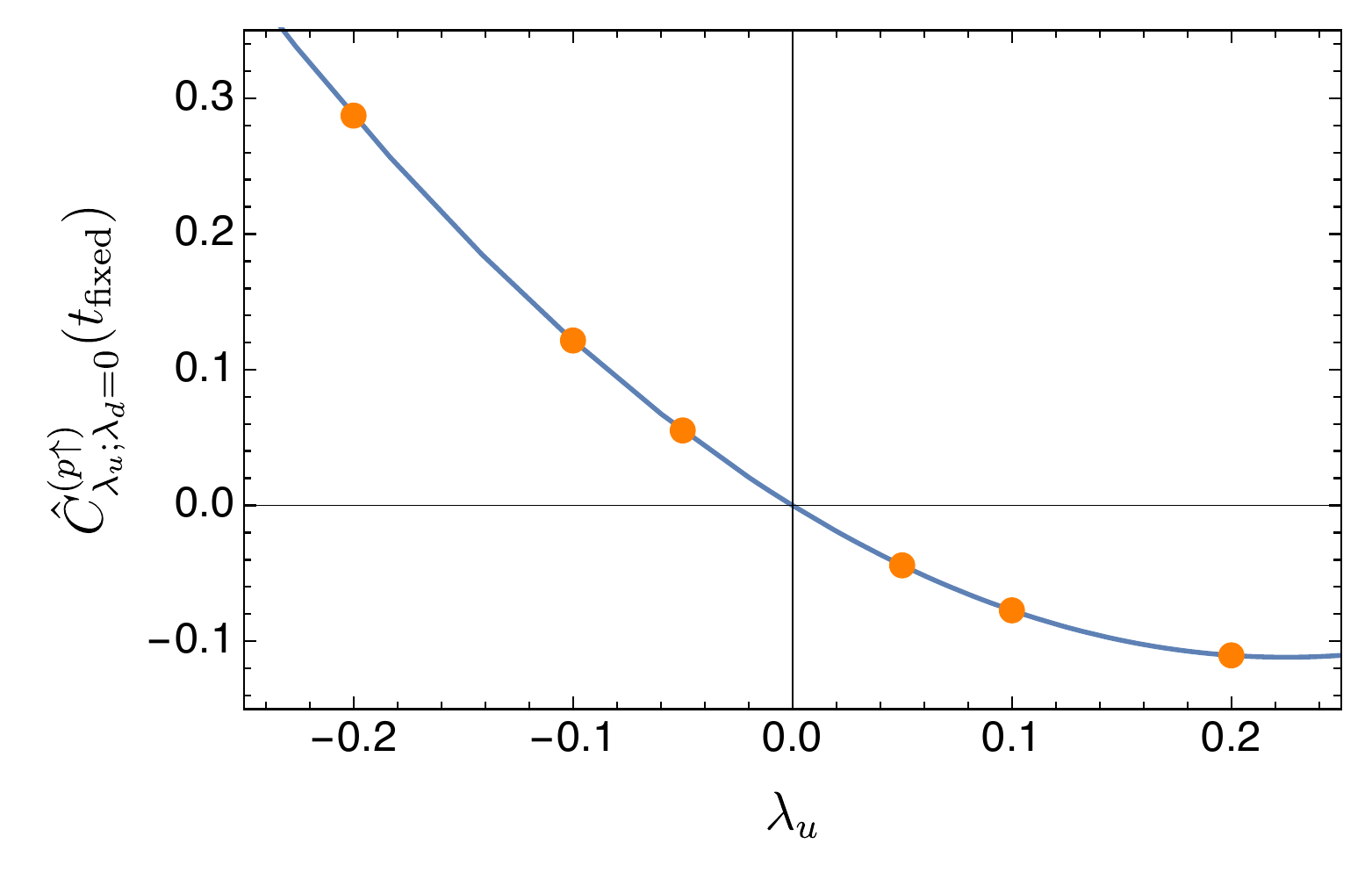}
		\label{fig:pufig}
	}
	\subfigure[]{
		\includegraphics[width=0.45\textwidth]{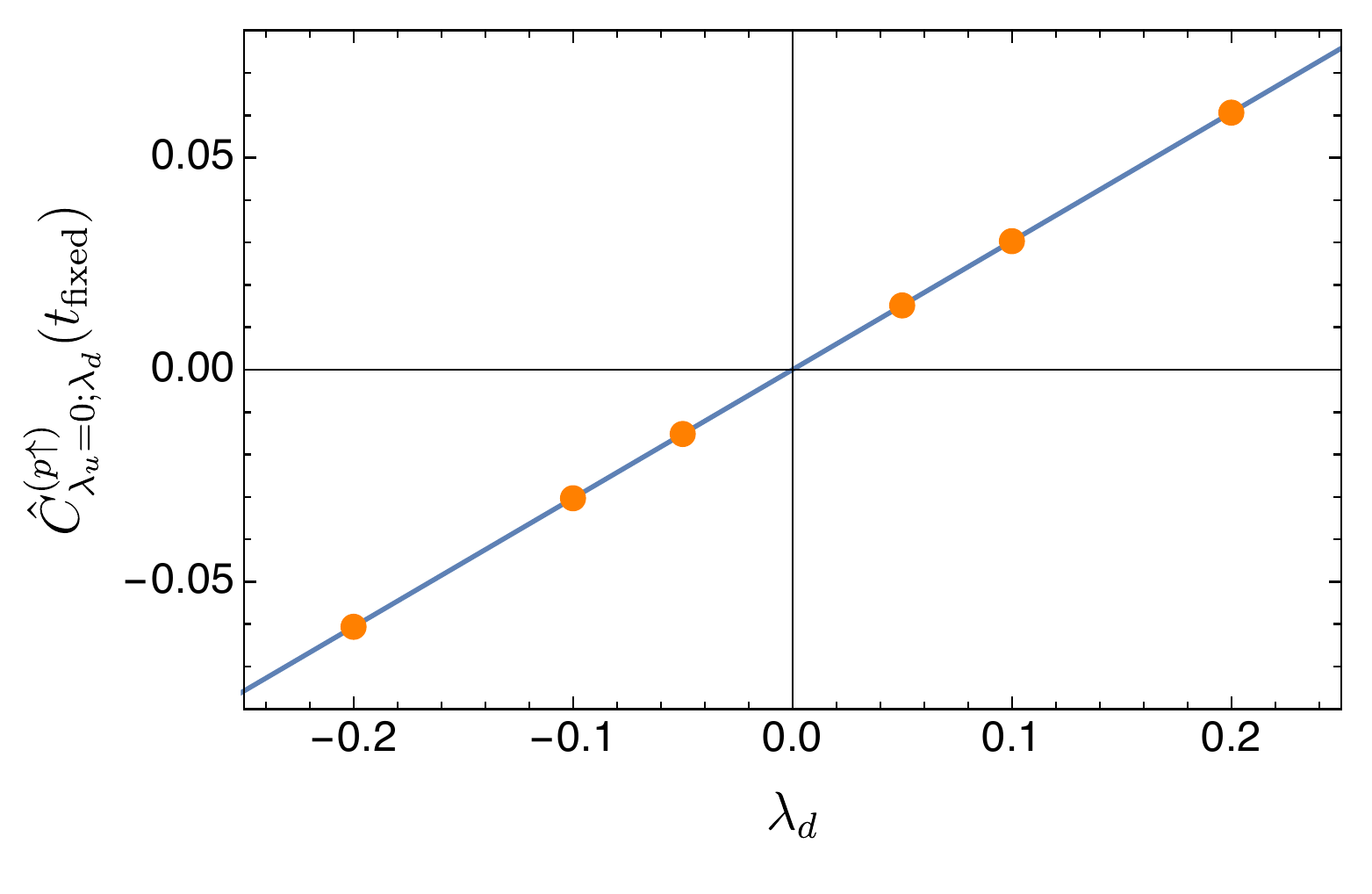}
		\label{fig:pdfig}
	}
	\subfigure[]{
		\includegraphics[width=0.45\textwidth]{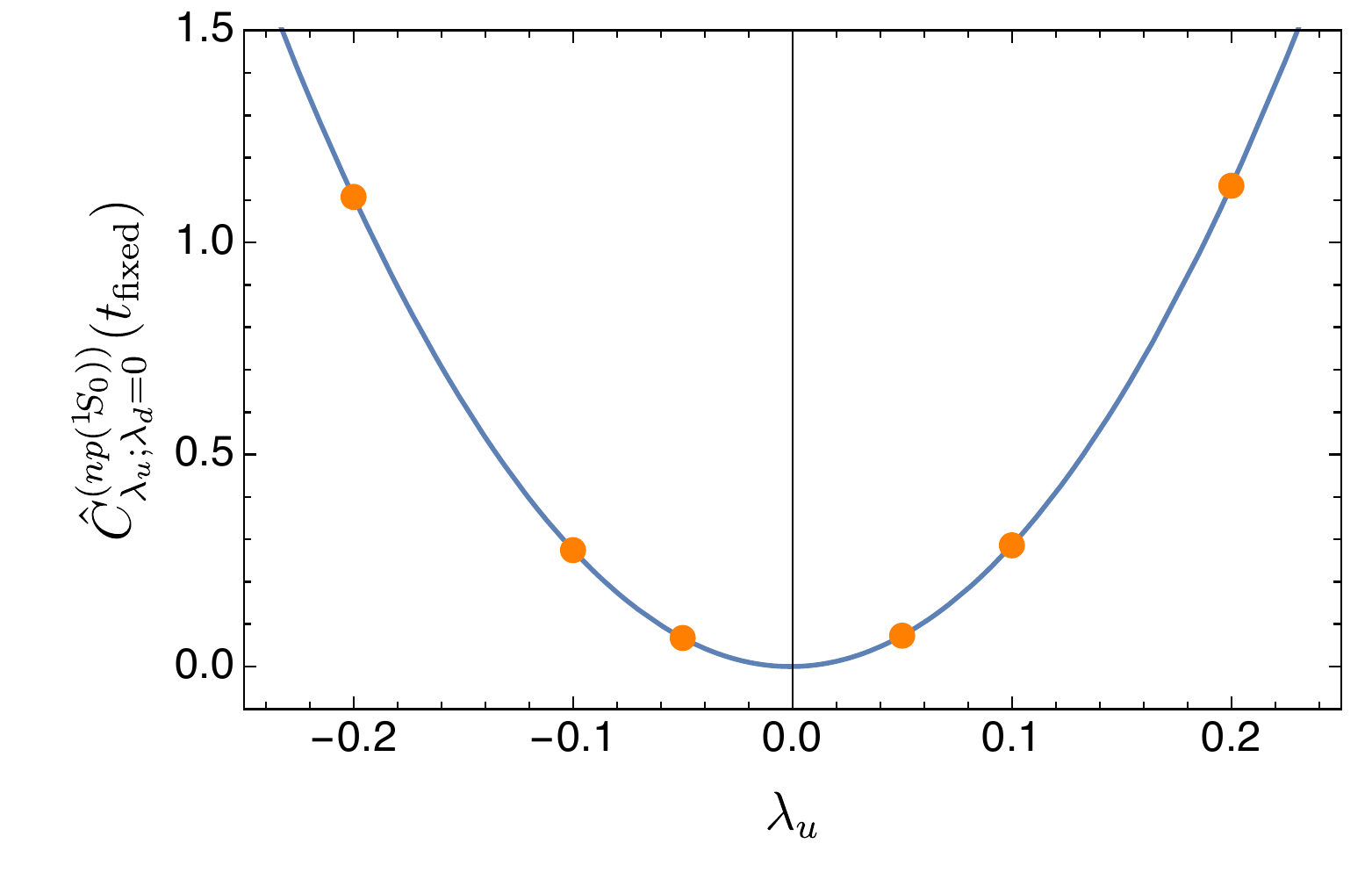}
		\label{fig:np1S0ufig}
	}
	\subfigure[]{
		\includegraphics[width=0.45\textwidth]{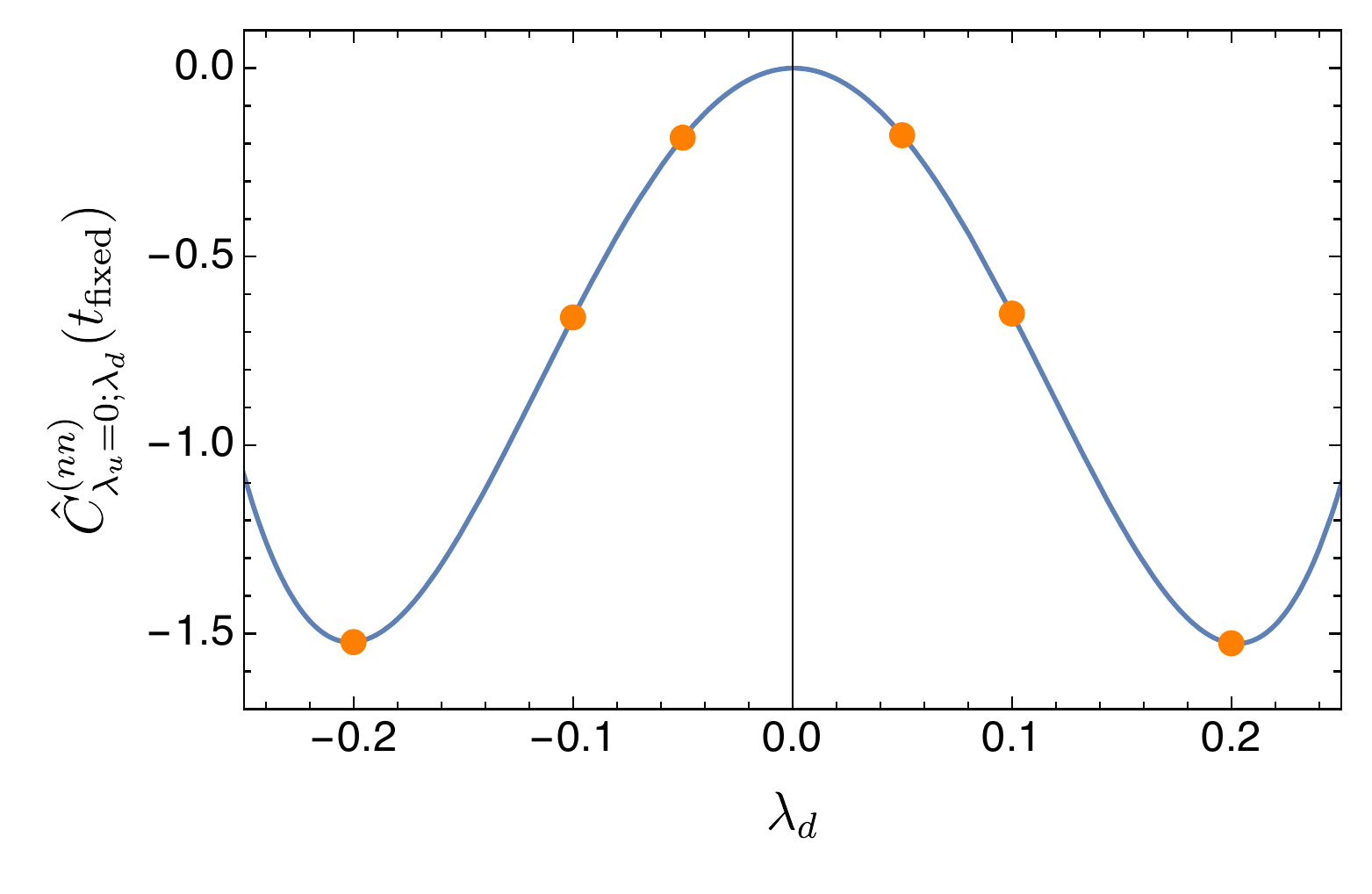}
		\label{fig:ddnfig}
	}
	\caption{
		The field-strength dependence of sample correlation functions constructed from compound propagators in the one-nucleon and two-nucleon sectors 
		evaluated on a given time slice on a given configuration. 
		The quantities shown are correlation functions with the zero-field limit subtracted.
		The polynomial fits (solid curves) are used to determine the requisite linear and quadratic responses.
	}
	\label{fig:field_response}
\end{figure}

In the large-time limit, the matrix element of interest can be uniquely isolated from the time dependence of 
the linear components of these correlation functions, 
for instance,
\begin{eqnarray}
\left.
C^{(p\uparrow)}_{\lambda_u;\lambda_d=0}(t)\right|_{{\cal O}(\lambda_u)} 
&= &
\sum_{t_1=0}^t \sum_{\frak{n},\frak{m}}
Z_\frak{n} Z^\dagger_\frak{m} e^{-E_\frak{n}(t-t_1)} e^{-E_\frak{m}t_1} \langle \frak{n}| \tilde{J}_3^{(u)} | \frak{m} \rangle 
\nonumber \\
&= &
\sum_{\frak{n},\frak{m}}
Z_\frak{n} Z^\dagger_\frak{m} \frac{e^{-E_\frak{n}t}- e^{-E_\frak{m} t}}{aE_\frak{m}-aE_\frak{n}} \langle \frak{n}| \tilde{J}_3^{(u)} | \frak{m} \rangle 
\nonumber \\
&\stackrel{t \to\infty}{\longrightarrow}&~
|Z_0|^2 e^{-E_0 t} \left[ c+
t\, \langle {p\uparrow}| {\tilde{J}_3^{(u)}} | {p\uparrow} \rangle +
{\cal O}(e^{-\hat{\delta} t})
\right],
\label{eq:Olambda2}
\end{eqnarray}
where  the  $Z_{\frak n}$ are proportional to the overlap of the interpolating operator 
onto a given state.  The energy splitting $\hat{\delta}$ corresponds to the gap between the ground state and excited states.

Extracting axial-current matrix elements in the two- and three-nucleon systems proceeds in a similar fashion. For the triton,
\begin{eqnarray}
C_{\lambda_u;\lambda_d=0}^{({}^3{\rm H}\uparrow)}(t)
& = &  
\sum_{\bf x}^{} 
\langle 0| \chi_{{}^3{\rm H}\uparrow}({\bf x},t) \chi^\dagger_{{}^3{\rm H}\uparrow}(0) |0 \rangle  
+ \lambda_u
\sum_{{\bf x},{\bf y}}\sum_{\tau=0}^t
\langle 0| \chi_{{{}^3{\rm H}\uparrow}}({\bf x},t) A_3^{(u)} ({\bf y},\tau)   \chi^\dagger_{{}^3{\rm H}\uparrow}(0) |0 \rangle 
\nonumber \\
&& \qquad+ e_2 \lambda_u^2 + e_3 \lambda_u^3 + e_4 \lambda_u^4, \qquad \\
C^{({}^3{\rm H}\uparrow)}_{\lambda_u=0;\lambda_d}(t) 
& =& 
\sum_{\bf x}
\langle 0| \chi_{{}^3{\rm H}\uparrow}({\bf x},t) \chi^\dagger_{{}^3{\rm H}\uparrow}(0) |0 \rangle  
+ \lambda_d 
\sum_{{\bf x},{\bf y}}\sum_{\tau=0}^t
\langle 0| \chi_{{}^3{\rm H}\uparrow}({\bf x},t) A_3^{(d)} ({\bf y},\tau) \chi^\dagger_{{}^3{\rm H}\uparrow}(0) |0 \rangle
\nonumber \\
&& \qquad+ f_2 \lambda_d^2 + f_3 \lambda_d^3 + f_4 \lambda_d^4 + f_5 \lambda_d^5,
\,
\end{eqnarray}
where the $e_i$ and $f_i$ are irrelevant terms. Similar expressions can be derived for the spin-down state.
In this case, data at four (five) values of the  
$u$ ($d$) field strength are required for direct extractions of the polynomial coefficients. Once the linear term is obtained, the remaining analysis required to determine the triton isovector axial matrix element proceeds as in Eq.~(\ref{eq:Olambda2}).

\subsection*{The Axial Current Renormalization Factor, $Z_A$}

For this ensemble of gauge-field configurations and the employed quark discretization, the renormalization factor, $Z_A$, is 
determined to be $Z_A=0.867(43)$.  
To determine this value, the matrix element of the vector-current operator was first calculated in the proton.  
As the proton has charge one, this determines the renormalization factor, $Z_V^{(p)}$, of the vector current 
to be 
$Z_V^{(p)}=0.86650(52)$~\footnote{The hadronic and quark superscripts are explicit for reasons of clarity.
} with high precision.
This is directly related to the renormalization constant at the quark level, $Z_V^{(q)}$, with $Z_V^{(p)}=Z_V^{(q)}$.
Up to lattice spacing artifacts, the tree-level renormalization constants for the axial-current and the vector-current 
operators are equal.   
Deviations will arise from finite lattice-spacing artifacts, which are expected to 
depend linearly on the lattice spacing, and from loop contributions involving the light quark masses at the scale of the lattice spacing.
The former are expected to be larger, and a conservative estimate is that they are at the $\sim 5\%$ level, which provides the uncertainty  associated with $Z_A=Z_A^{(q)}=Z_V^{(q)} + {\cal O}\left(a\right)$. 

The renormalization, $Z_A$, has been precisely determined at a  pion mass of $m_\pi\sim 317~{\rm MeV}$
with the same quark discretization and analogous gauge-field configurations, but employing non-perturbative renormalization techniques~\cite{GreenLatt2016,Yoon:2016jzj}.  
A value of  $Z_A=0.8623(01)(71)$ was obtained, that is consistent with the value  obtained here at the 
heavier pion mass.

\subsection*{Effective Masses for the Nucleon and Light Nuclei}

The effective masses of the nucleon, deuteron, dineutron and the triton, obtained in these calculations using two different sink structures, 
are shown in Fig.~\ref{fig:allEMPs}.\footnote{
	The present calculations have been accomplished with many fewer source locations than used in 
	Ref.~\cite{Beane:2012vq} and, as a result, the statistical precision of the  effective masses in this work is comparatively less.
}
\begin{figure}[!ht]
	\includegraphics[width=0.95\textwidth]{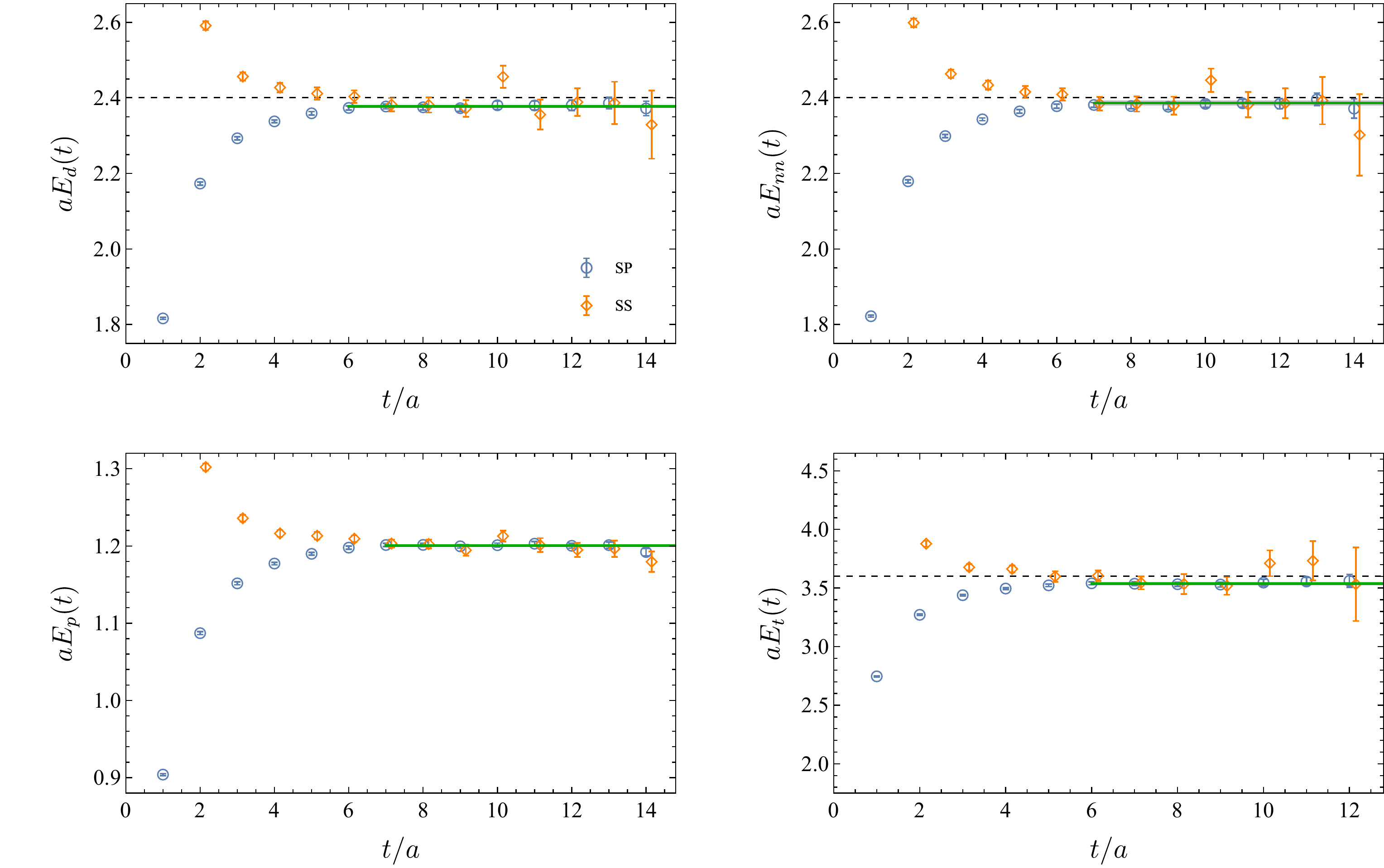}
	\caption{
		The  effective mass plots for the deuteron (top-left), dineutron (top-right), nucleon (bottom-left) and the triton (bottom-right).  
		Energies extracted from fits to the plateau regions, starting at time slices t=6 or 7, agree with the 
		previous determinations~\cite{Beane:2012vq} within uncertainties.
	}
	\label{fig:allEMPs}
\end{figure}
For the nucleon and each of the nuclei, the effective masses reach plateau values (within uncertainties) by time slices t=6 or 7, 
and are consistent with those extracted from more precise calculations~\cite{Beane:2012vq} within the uncertainties of both calculations.
As such, ratios of these correlation functions can be used to extract the  differences in ground-state energies starting at t=6 or 7
with only exponentially small  excited state contamination.

In Ref.~\cite{Iritani:2017rlk}, Iritani {\it et al.} argue that all calculations in the literature showing evidence for bound deuteron and di-nucleon states at this quark mass are erroneous primarily because they claim that  the extracted finite-volume energies are dependent on the source and sink interpolators that are used. In Fig.~\ref{fig:binding}, all extractions of the ground-state energies of the ${}^3S_1$ and ${}^1S_0$ two-nucleon systems in three different volumes are shown along with simple fits to the relevant sets. These clearly show the statistical consistency of the different extractions, invalidating the claims of Ref.~\cite{Iritani:2017rlk}. Further discussions demonstrating the invalidity of the claims in Ref.~\cite{Iritani:2017rlk}  can be found in Ref.~\cite{comment}. 
\begin{figure}[th!]
	\includegraphics[width=0.95\columnwidth]{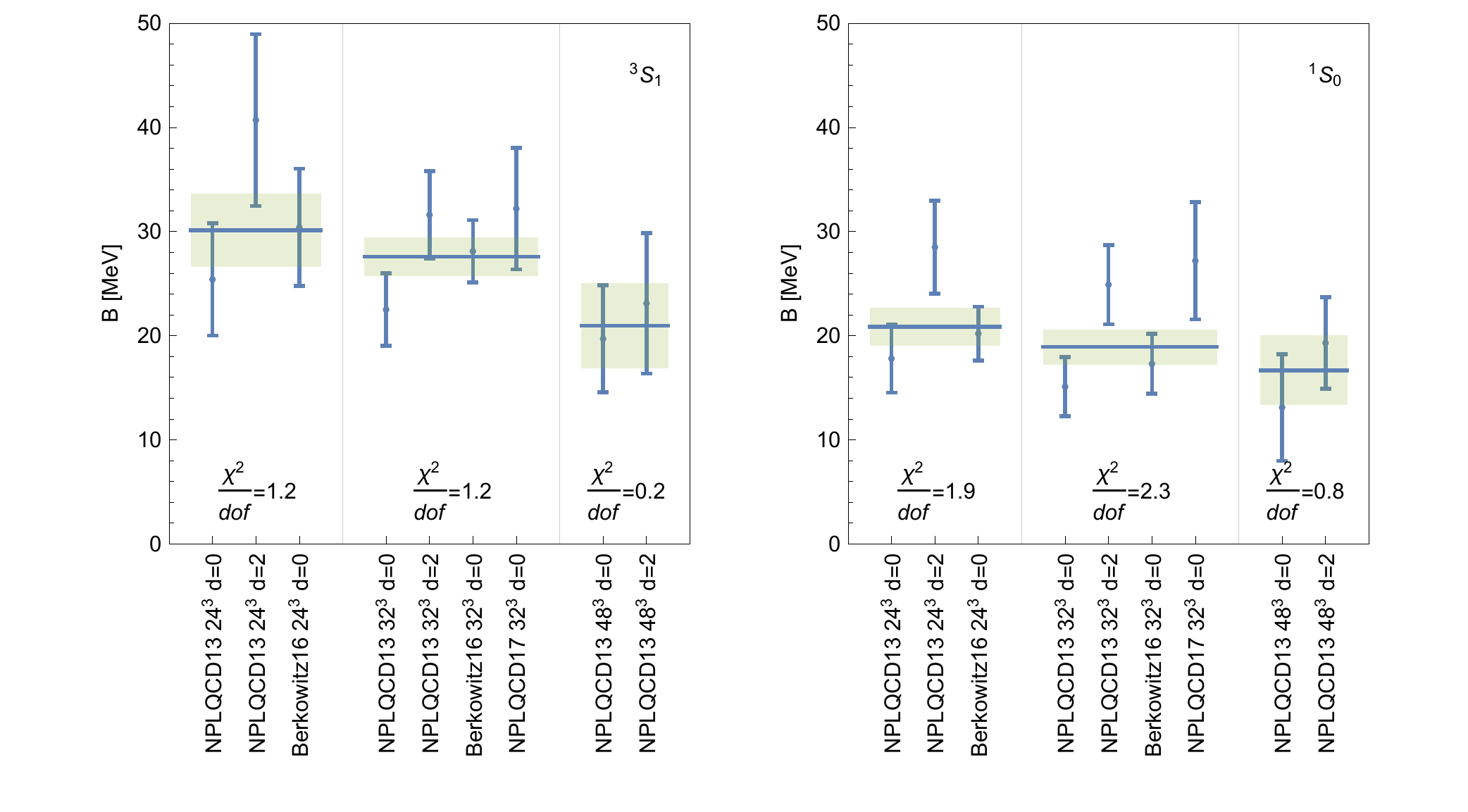}
	\caption[.]{Binding energies for the $\siii$ and $\si$ ground states at $m_\pi= 806$ MeV found in the literature: NPLQCD13 \cite{Beane:2012vq}, Berkowitz16 \cite{Berkowitz:2015eaa}, and NPLQCD17 \cite{Tiburzi:2017iux} ($d=0$ and $d=2$ refer to the magnitude of the centre-of-mass momentum used in the calculations in units of $2\pi/L$). The three regions in each panel correspond to three different volumes: $L=24$, 32, and 48 from left to right. Uncertainties listed in the original references are combined in quadrature. The horizontal lines and shaded bands represent the central value and one standard deviation bands on fits to the indicated data, respectively.}	\label{fig:binding}
\end{figure}

\subsection*{Uncertainty propagation in EFT matching}

In propagating the uncertainty on the extracted lattice value of $L_{1,A}^{sd-2b}$  through
to the determination of $\Lambda(0)$ and $L_{1,A}$ using Eq.~(8) and (14), the experimental values of the following quantities are used:
\begin{eqnarray*}
	g_A=1.2579(45), \\
	a_{pp}=-7.8063(29)\ {\tt fm}, \\
	r_1=2.79(6)\ {\tt fm},  \\
	\rho=1.764(2)\ {\tt fm}, \\
	\gamma=45.86(8)\ {\tt MeV}.
\end{eqnarray*}
The ranges shown above are sampled assuming normal distributions and their effect is included as part of the systematic uncertainty on the values of $\Lambda(0)$  and $L_{1,A}$  presented in the main text.

In addition, the following relationship between $L_{1,A}$ and $\overline{L}_{1,A}$ 
is used:
\begin{eqnarray}
\overline{L}_{1,A}= \frac{4\pi}{g_A M^2}
\left[ \frac{L_{1,A}}{C^{{\left(\siii\right)}}_0 C^{{\left(\si\right)}}_0} - \frac{g_AM}{2}\frac{(C^{{\left(\siii\right)}}_2+C^{{\left(\siii\right)}}_2)}{C^{{\left(\siii\right)}}_0 C^{{\left(\si\right)}}_0} \right],
\label{eq:L1Abar}
\end{eqnarray}
where $C_i^{(\si)}$ and $C_i^{(\siii)}$ are low-energy constants of the two-nucleon Lagrangian \cite{Chen:1999tn,ButlerChen}
\begin{eqnarray*}
	C_0^{(\siii)}(\mu) &=& -\frac{4\pi}{M}\frac{1}{\mu-\gamma}, \\
	C_2^{(\siii)}(\mu) &=& \frac{2\pi}{M}\frac{\rho}{(\mu-\gamma)^2}, \\
	C_0^{(\si)}(\mu) &=& \frac{4\pi}{M}\left[{1\over a_{pp}} -\mu+\alpha M\left( \ln\frac{\mu\sqrt{\pi}}{\alpha M}+1 -{3\over 2} \gamma_E\right)
	\right]^{-1},
	\\
	C_2^{(\si)}(\mu)&=&{M\over 8\pi} r_1 \left[C_0^{(\si)}(\mu)\right]^2,
\end{eqnarray*}
where $\gamma_E$ is the Euler-Mascheroni constant. 
The value of $L_{1,A}$ is obtained at a renormalization scale $\mu=m_{\pi}$.

\subsection*{Quark-mass dependence}
In the low-energy effective field theory, the operator structure of the two-nucleon correlated interaction with the isovector 
axial-vector field is of the same form as that of the two-nucleon correlated interaction with the magnetic field, described by 
the short-distance counterterm $\overline{L}_1$.
Ref.~\cite{Beane:2015yha}  presented lattice 
QCD calculations of 
$\overline{L}_1$ at two unphysical quark masses, and showed that simple polynomial extrapolations to the physical values of the quark masses were consistent with the 
experimentally extracted value, and, surprisingly, were independent of the quark masses within uncertainties.
In the current work, the analogous short-distance counterterm $\overline L_{1,A}$ associated with $pp$ fusion has been identified.
It is  expected 
(guided by the behavior of $\overline{L}_1$) 
that it too exhibits mild 
variation with the pion mass down to the physical point.
A systematic uncertainty has been included to account for deviations from this expectation as discussed in the main text. 
Further calculations, at lighter pion masses, along with smaller lattice spacings and larger spacetime volumes,
are required to reduce the size of this systematic uncertainty.

Close to the chiral limit, neither the pionless EFT analysis nor the lattice QCD calculations can be used.
Lattice QCD results at and near the chiral limit will have a quite different structure due to the finite volume:
correlation functions will typically have power-law scaling for volumes that are small compared with the inverse pion mass; massless excitations will eliminate gaps in the spectrum; and thermal effects from backward propagating pions will be 
problematic.
Furthermore, a pionless EFT will not be a useful tool as the energy regime for which it will be applicable is set by the pion mass.  Pionful EFTs will be required for matching and, because of the finite volume, the $\epsilon$-regime power-counting will be appropriate, requiring integration over the 
zero-mode coset-space.
At the physical quark masses, these complexities will be manageable.

\end{widetext}
\end{document}